\documentclass[aps,prl,twocolumn,groupedaddress,showpacs]{revtex4}

\usepackage{graphics}
\usepackage{epsfig}
\usepackage{amsmath}
\usepackage{amscd}
\usepackage{amssymb}
\usepackage{amsthm}
\usepackage{amsfonts}
\usepackage{amstext}
\usepackage{textcomp}
\usepackage[dvips]{color}

\begin{document}


\title{Quantum-Well Wavefunction Localization and the Electron-Phonon Interaction in Thin Ag Nanofilms}


\author{S. Mathias,$^1$ M. Wiesenmayer,$^1$ M. Aeschlimann,$^1$ M. Bauer$^2$}
\affiliation{$^1$Department of Physics, University of
Kaiserslautern, 67663 Kaiserslautern, Germany\\$^2$Institut f\"{u}r
Experimentelle und Angewandte Physik, Universit\"{a}t Kiel, 24098
Kiel, Germany}


\date{\today}

\begin{abstract}
The electron-phonon interaction in thin Ag-nanofilms epitaxially
grown on Cu(111) is investigated by temperature-dependent and
angle-resolved photoemission from silver quantum-well states. Clear
oscillations in the electron-phonon coupling parameter as a function
of the silver film thickness are observed. Different from other thin
film systems where quantum oscillations are related to the
Fermi-level crossing of quantum-well states, we can identify a new
mechanism behind these oscillations, based on the wavefunction
localization of the quantum-well states in the film.

\end{abstract}

\pacs{73.50.Gr, 73.21.-b, 71.18.+y, 79.60.Dp}

\maketitle

\begin{figure}
\includegraphics[width=220pt,keepaspectratio]{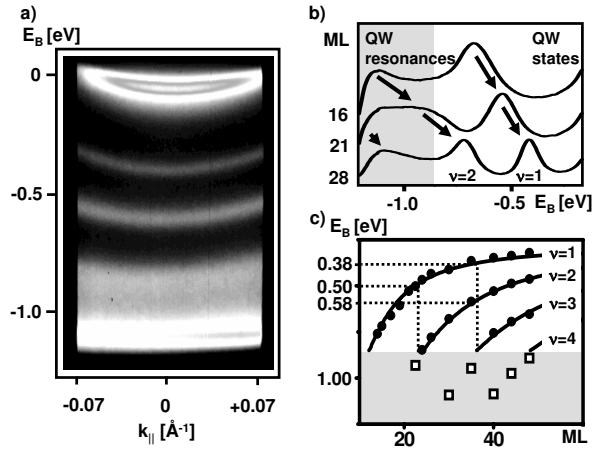}
\caption{(a) Typical 2D photoemission map for 36$\,$ML Ag/Cu(111),
$T=211\,$K, $hv=6\,$eV. (b) Transition of a QW-resonance into a
QW-state with increasing Ag-thickness. The shaded area indicates the
projected Cu bulk bands to the surface. (c) Energies of the $\nu$=1
to $\nu$=4 QW-states (dots) and QW-resonances (squares) as a
function of nominal film thickness \cite{mathias}. The solid curves
are calculations using the phase accumulation model with
$m^*/m$=0.65, in agreement with data published before in
\cite{chiangagcubloch}. The dotted lines link the QW-state
QW-resonance transitions occurring at 23 ML and 35 ML to the binding
energy scale.}
 \label{keilmitfit}
\end{figure}
\begin{figure}
\includegraphics[width=220pt,keepaspectratio]{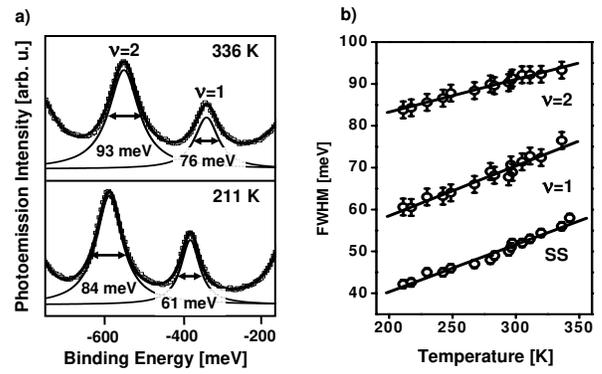}
\caption{a) Temperature dependent photoemission spectra of 36 ML
Ag/Cu(111) for 336 K and 211 K. The $\nu$=1 and $\nu$=2 peaks shift
in binding energy and the linewidth increases with increasing
temperature. b) Linewidths (dots) and linear fit (solid line) of the
QW-states and the Shockley surface state as a function of
temperature for this film.
 \label{temperature}}
\end{figure}
\begin{figure}
\includegraphics[width=220pt,keepaspectratio]{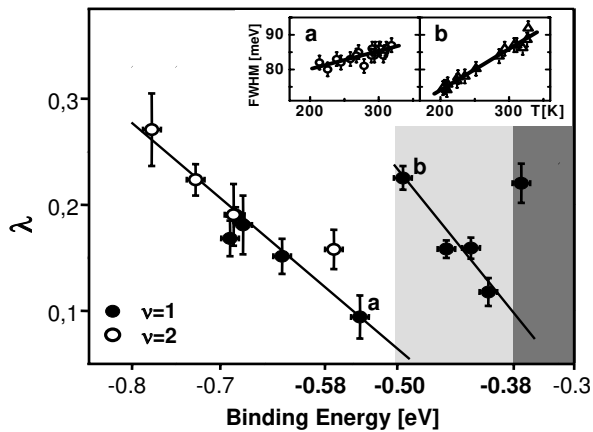}
\caption{Electron-phonon coupling parameters for the QW-states
$\nu$=1 (full circles) and $\nu$=2 (empty circles) as a function of
binding energy, deduced from T-dependent linewidth measurements as
shown in the inset and Figure \ref{temperature}b. Oscillations in
$\lambda$ occur at binding energies (bold printed) at which a
QW-resonance to QW-state transition occurs, as can be seen in Figure
1c. For reference, the gray shaded areas mark the binding energy
regimes of increased electron-phonon scattering due the $\nu$=1 to
$\nu$=2 QW-state interaction (light gray) and due to the $\nu$=1 to
$\nu$=3 QW-state interaction (dark gray). The inset shows the
temperature dependence of the linewidth of the $\nu$=1 QW-state
right before (a) and after (b) the $\lambda$-step at 0.5 eV binding
energy.}
 \label{lambdabinding}
\end{figure}
\begin{figure}[t]
\includegraphics[width=240pt,keepaspectratio]{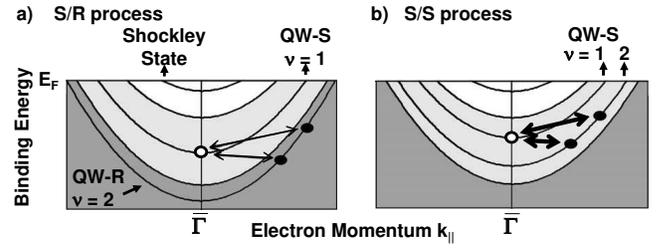}
\caption{Schematic illustration of the QW-state hole relaxation
paths. Hole states are indicated by open circles, while electrons
are indicated by solid circles. The dark shaded area is the
projected bulk bands of the Cu(111)-substrate to the surface. a)
Phonon mediated scattering into the delocalized $\nu$=2
QW-resonance; b) enhanced electron-phonon coupling due to
phonon-mediated interband scattering into the localized $\nu$=2
QW-state. The energy differences are highly exaggerated.
 \label{relaxschema}}
\end{figure}

Quantum oscillations of various physical properties in thin film
quantum-well systems, which are controlled by the thickness of the
overlayer film, have generated constant interest in recent years. A
very prominent example are quantum oscillations in the
superconducting transition temperature of thin Pb-films with varying
thicknesses \cite{guosuper,eomsuper,oezersuper,chiangtc}. In the
same context, quantum oscillations in the electron-phonon coupling
have been observed in, for example, Ag and Pb
\cite{luhosci,vallaosci}. Additional examples include oscillations
in the magnetic coupling \cite{parkinmagnetic}, the stability of
thin films \cite{luhstability}, work-function
\cite{paggelworkfunction} and electronic growth
\cite{zhangelectronic}. All these systems are characterized by the
existence of discrete electronic quantum-well states (QW-states)
and, up till now, any observed oscillatory behaviour has been
related to a Fermi-level crossing of a QW-state at distinct film
thicknesses. In this paper, we give first evidence for a different
mechanism in thin film systems capable of promoting quantum
oscillations. We show that quantum oscillations in the
electron-phonon coupling of ultrathin Ag-films on Cu(111) are driven
by transitions of a QW-state into a QW-resonance at distinct film
thicknesses. This localization process enhances the probability of
electron momentum transfer processes and, therefore, promotes
electron-phonon scattering. In analogy to the relevance of
Fermi-level crossing of QW-states we also expect this kind of
wavefunction localization to alter a variety of relevant material
properties in a similar quantum
oscillatory way.\\
Our photoemission data were recorded at normal emission using a
hemispherical energy analyzer for parallel energy and momentum
detection ($\Delta$E$<$20 meV, $\Delta\phi<$ 0.15\textdegree).
Frequency-quadrupled UV light (h$\nu$=6eV) of the output of a small
bandwidth ($h\nu$=1.5 eV) pulsed Ti:Sapphire laser system was
focused onto the surface to a spotsize of approximately 150 $\mu$m.
Silver films were grown at room temperature followed by short flash
of the sample to 600 K \cite{mathias}. Figure \ref{keilmitfit}a
displays a $E(k_{||})$-Photoemission map of a 36 ML thick silver
film recorded at a sample temperature of 211 K (intensity in gray
scale as a function of energy and parallel momentum). The spectrum
shows an intense signal from the Shockley surface-state just below
the Fermi-level at a binding energy of about 40 meV. The spectral
features at binding energies of 390 meV, 600 meV and 840 meV are due
to the  $\nu$=1, $\nu$=2 and $\nu$=3 QW-states localized within the
silver film overlayer \cite{chiangagcubloch,tanaka}. The appearance
of these states is possible due to the presence of the sp band-gap
of the Cu(111) substrate, which extends from -4.25 eV to 0.85 eV
binding energy at $k_{||}$=0\,${\AA}^{-1}$. Figure \ref{keilmitfit}b
shows a set of three photoemission spectra at normal emission for 16
ML, 21 ML and 28 ML of Ag on Cu(111). In agreement with previous
reports \cite{chiangagcubloch,tanaka}, we observe a shift of the
QW-states to lower binding energies with increasing silver coverage
asymptotically approaching the energy of the upper band edge of the
silver sp band (300 meV). Furthermore, additional spectral features
significantly below the lower edge of the sp band-gap (850 meV) are
visible (gray shaded area). These are due to photoemission from
quantum-well resonances (QW-resonances) in the silver film. These
types of states are supported by a finite electron reflectivity at
the silver/copper interface, which is due to wavefunction mismatch
between the two materials and, in the present system, is enhanced by
a lattice mismatch of about 13\% between silver and copper
\cite{meunier}. In contrast to pure QW-states, these states are
obviously in resonance with the copper sp-band and couple to real
electron states within the copper substrate. The coupling gives rise
to extended penetration (delocalization) of these states into the
copper going along with a reduction of the wavefunction amplitude
within the silver film \cite{woodruffueber,chiangueber}. Figure
\ref{keilmitfit}c shows results from a quantitative analysis of
photoemission maps recorded at varying film thicknesses. Monotonous
transitions of the $\nu$=2, $\nu$=3 and $\nu$=4 QW-resonances (open
squares) into QW-states occur at film thicknesses of about 23 ML, 35
ML and 47 ML. Additional evidence for the change in the character of
these states at these film thicknesses is a narrowing of their
linewidths right at these critical thicknesses in agreement with
reference \cite{chiangueber}.\\
A calculation of the binding energy of the  $\nu$=1 to $\nu$=4
QW-states was performed using the phase accumulation model following
the details described in references \cite{smithmodel} and
\cite{paggeltwoband} (see solid lines in figure \ref{keilmitfit}).
The model reproduces our experimental data perfectly for
$m^*/m$=0.65 using an interfacial phase contribution given by
\cite{smithmodel,uptonthermal}
\begin{equation}
\Phi_{C}(E)=2\arcsin{\sqrt{(E-E_L)/(E_U-E_L)}},\label{phic}
\end{equation}
where $E_L=-0.85\,$eV and $E_U=4.25\,$eV are the lower and upper
edge of the Cu(111) band gap \cite{paniagodata}, respectively. The
linewidth $\Delta E$ of the QW-state peaks is related to the
lifetime of the photohole via the uncertainty principle and is
governed to a considerable extent by electron-phonon interactions.
An increased phonon population at elevated temperatures leads to an
enhanced phonon scattering rate, and hence a shorter lifetime and an
increase in the QW-state linewidth. Figure \ref{temperature}a shows
photoemission spectra of the $\nu$=1 and $\nu$=2 QW-states for a 36
ML thick silver film at temperatures of 211 K and 336 K (open
squares). The two Lorentzians result from fits of the QW-states
while the solid lines represents the total fit of the spectra.
Figure \ref{temperature}b summarizes for this film the linewidths of
the $\nu$=1 and $\nu$=2 QW-state and the Shockley surface state as a
function of temperature between 200 and 350 K. An important
parameter that can be directly deduced from these data is the
electron-phonon coupling parameter $\lambda$ \cite{grimvall}, which
is given by
    \begin{equation}
    \lambda=\frac{1}{2\pi k_B}\frac{d(\Delta E)}{dT}. \label{formellambda}
    \end{equation}\\
$d(\Delta E)/dT$ is the slope of the observed linear temperature
dependence. The electron-phonon coupling parameter clearly depends
on the quantum
number $\nu$ of the QW-states.\\
Next, Figure \ref{lambdabinding} shows electron-phonon coupling data
obtained from QW-state linewidth measurements for silver films of 15
ML to 36 ML thickness. The data are displayed here as a function of
the QW-state binding energy. Full and open circles correspond to the
values obtained for the  $\nu$=1 and $\nu$=2 QW-states,
respectively. The general trend observed is an overall decrease of
$\lambda$ with decreasing binding energy (solid lines). This trend
is, however, interrupted by steps which are most pronounced at
QW-state binding energies of about 0.50 eV and 0.38 eV for the
$\nu$=1 state, but are also visible at a binding energy of 0.58 eV
for the $\nu$=2 state.\\
The linear decrease is consistent with a simple model previously
used to quantitatively describe results for the Ag/Fe(100) system
\cite{luhosci}. This model allows us to assign the binding-energy
dependence of $\lambda$ to the phase shift variation of $\pi$ across
the confinement gap at the silver/copper interface \cite{smithmodel}
(see equation (\ref{phic})). This phase shift $\Phi_{C}(E)$
modulates the amplitude of the QW-state's wavefunction at the
interface and, therefore, affects the coupling efficiency of the
QW-states to interfacial phonons. Since this modulation is only
dependent on the binding energy of the QW-states, the behavior
within our film thicknesses is independent of the quantum number $\nu$.\\
However, the experimentally observed deviations from this linear
trend at 0.50 eV and 0.38 eV binding energy ($\nu$=1) and  at 0.58
eV binding energy ($\nu$=2) cannot be reproduced by this model. The
observed step-like behavior requires a corresponding change in the
interfacial phase contribution, which
is not existent for the investigated Ag/Cu system.\\
Oscillations in the electron-phonon coupling parameter as a function
of quantum-well thickness have been observed for other quantum-well
systems before \cite{guosuper,eomsuper,oezersuper,luhosci,
vallaosci,parkinmagnetic,luhstability,paggelworkfunction,zhangelectronic}.
In all these cases, the origin of the oscillations was related to a
Fermi-level crossing of a QW-state. The driving force for this
effect is the strong modulation of the occupied density of states
near $E_F$ as the QW-states cross the Fermi-level with increasing
film thickness. However, for the Ag/Cu(111) system, this scenario
can be excluded. At $k_{||}=0\,{\AA}^{-1}$ the QW-states
asymptotically converge at the lower edge of the silver sp band-gap
at 300 meV binding energy, which is well below the Fermi-edge (see
Figure \ref{keilmitfit}). Therefore, a Fermi-level crossing is not
possible at all film-thicknesses and consequently cannot cause the
observed oscillations in $\lambda$.\\
We found instead a direct correlation between the appearance of
these oscillations and the critical film thicknesses where the
transitions from the $\nu$=2 and $\nu$=3 QW-resonances into the
associated QW-states occur. For instance, the discontinuity in the
$\nu$=1 QW-state at 0.5 eV binding energy is seen at a silver film
thickness of 23 ML, exactly where we observe the transition of the
$\nu$=2 QW-resonance into the $\nu$=2 QW-state (see dotted lines in
Figure \ref{keilmitfit}c). In addition, both the second step of the
$\nu$=1 QW-state at 0.38 eV binding energy and the step seen in the
$\nu$=2 QW-state at 0.58 eV binding energy correlate with the
$\nu$=3 QW-resonance transition into the $\nu$=3 QW-state at 35 ML.
We therefore conclude that the origin of the oscillations in the
electron-phonon coupling parameter $\lambda$ for the Ag/Cu(111)
system are directly related to these changes in the electronic
structure, which occur at specific film-thicknesses as a
QW-resonance undergoes the transition into a QW-state.\\
The electron phonon-interaction in the silver film is governed by
the available decay channels, the available phase space for each
decay channel, and the matrix elements for the scattering processes
\cite{echeniqueueber,grimvall,paggeldband}. The electron-phonon
scattering process responsible for the phonon-mediated hole
relaxation probed in the photoemission experiment requires, in
particular, a source of electron momentum to compensate for the
momentum of the absorbed or created phonon. Highly efficient in this
means are QW-states interband scattering processes between two
QW-states of different quantum number $\nu$ (see Figure
\ref{relaxschema}b). This has been previously shown, for example, in
the case of d-band QW-states relaxation \cite{paggeldband} and, in a
similar manner, in connection with the decay of image potential
states at Cu(100) surfaces \cite{berthold}. An important
contribution to the matrix element of such a process is the
wavefunction overlap between the two involved QW-states.\\
In this context, we now consider in more detail the potential decay
channels available for the $\nu$=1 hole created by the photoemission
process (see Figure \ref{relaxschema}). Three different
phonon-mediated scattering processes can be discriminated: the
interaction of the $\nu$=1 QW-state hole with (i) the Cu bulk-bands
and, depending on the silver film thickness, the interaction with
(ii) the QW-resonance (S/R process, see Figure \ref{relaxschema}a))
and the interaction with (iii) the $\nu$=2 QW-state (S/S process,
see Figure \ref{relaxschema}b). Due to the negligible wavefunction
overlap between the $\nu$=1 QW-state confined in the silver film and
the bulk states of the copper substrate, the contribution of process
(i) to the electron-phonon cross-section is of negligible
importance. However, for processes (ii) and (iii), both the initial
hole state ($\nu$=1 QW-state) and final hole state ($\nu$=2
QW-resonance and QW-state, respectively) exhibit a considerable
wavefunction amplitude within the silver film that promotes the
electron-phonon interaction. However, the leakage of the
QW-resonance into resonant Cu-bulk states notably reduces its
wavefunction amplitude in the silver film in comparison to the
$\nu$=2 QW-state. Indicative of this leakage are the low intensity
and the large linewidth $\Delta E$ of the $\nu$=2 QW-resonance above
a value of 400 meV at 300 K (Figure \ref{keilmitfit}b). From this
point of view, the phonon-mediated S/R scattering process must be
less efficient than the S/S process and this enables us to
understand the microscopic mechanism behind the step-like increase
in $\lambda$ at the critical film thickness of 23 ML. For a silver
film just below this thickness, the electron-phonon coupling is
governed by a phonon-mediated S/R scattering process. As the film
thickness increases, the $\nu$=2 QW-resonance is transformed into
the $\nu$=2 QW-state corresponding to a wavefunction localization
and an increase in the wavefunction amplitude. Right above the
critical film thickness, the electron-phonon coupling is therefore
mediated by the S/S scattering process, which is characterized by an
enhanced interaction probability as observed in the experiment. The
oscillatory behavior in $\lambda$ is therefore expected at a
periodicity given by the rate at which a QW-resonance into QW-state
transition takes place, in perfect agreement with our observations
at 23 ML and 35 ML film thickness. This argument is also valid for
the discontinuity of the $\nu$=2 QW-state at 580 meV binding energy
corresponding to the $\nu$=3 QW-resonance QW-state transition at 35
ML thickness.\\
The ratio of the efficiency between the S/R scattering process and
the S/S process can be quantified by an estimate of the wavefunction
amplitude of the $\nu$=2 QW-resonance relative to the $\nu$=2
QW-state in the silver film. For a finite reflection coefficient
$R_{\text{QWR}}$ ($R_{\text{QWR}}<1$) of the QW-resonance at the
silver/copper interface, the ratio of the wavefunction amplitude
between the QW-resonance and QW-state ($R_{\text{QWS}}\approx$1) is
given by $R_{\text{QWR}}/R_{\text{QWS}}\approx R_{\text{QWR}}$.
$R_{\text{QWR}}$ can be deduced from the measured linewidth $\Delta
E$ of the QW-resonance. According to reference
\cite{paggelinterferometer}, $\Delta E$ depends on the number of
monolayers N of the silver film, $R_{\text{QWR}}$ and the mean free
path L of the electrons,
\begin{equation}
\Delta E=\Gamma \eta
\frac{1-R_\text{QWR}\,exp({-1/\eta})}{R_\text{QWR}^{1/2}exp({-1/(2\eta))}}
\end{equation}
with $\eta=L/(Nt)$, t is the monolayer thickness and $\Gamma$ is the
intrinsic linewidth of the QW-resonance. For $L\approx Nt/2$
\cite{ashcroft} and $\Gamma$=90 meV as deduced from an extrapolation
of the intrinsic linewidth data of the QW-state into the
QW-resonance regime we obtain a value of the reflectivity
$R_{\text{QWR}}<$1/4. This means that the wavefunction amplitude of
the QW-resonance is reduced in comparison to the wavefunction
amplitude of the QW-state by a factor of about 4. Also of the same
order of magnitude is the ratio of the wavefunction overlap between
the S/R and S/S scattering processes and, consequently, the expected
ratio in the electron-phonon coupling constant. In this experiment,
we find the ratio for the $\nu$=1 QW-state is equal to a value of
about 3.5(1). Taking into account that the above estimation
exclusively focuses on the effect of the wavefunction amplitude, the
experimental and calculated values are
in reasonable agreement.\\
In conclusion, we have observed oscillations in the electron-phonon
coupling parameter in thin Ag overlayer films on Cu(111). We have
shown that these oscillations correlate with the transition of a
QW-resonance into a QW-state in the silver film. Furthermore, we
attribute the microscopic mechanism for these changes to the
localization of the QW-state wavefunction within the silver film,
which is going along with such a transition. In addition to
quantum-oscillations in the electron-phonon coupling parameter, we
propose that this mechanism is of more general relevance for other
thin film properties similar to the effect of the QW-state
Fermi-level crossing. For example, one property for that oscillatory
behavior has been shown in systems with oscillating $\lambda$ is
superconductivity
\cite{guosuper,eomsuper,oezersuper,chiangtc,zhangphonon}.

\begin{acknowledgments}
Special thanks go to M. Wessendorf.
This work was supported by the DFG
through SPP 1093 and the Stiftung Innovation RLP.
\end{acknowledgments}


\end{document}